\documentclass[10pt,letterpaper]{article}
\usepackage[top=0.85in,footskip=0.75in]{geometry}
\usepackage{verbatim}

\usepackage{amsmath,amssymb}

\usepackage{numprint}

\usepackage{changepage}

\usepackage[utf8x]{inputenc}

\usepackage{textcomp,marvosym}

\usepackage{cite}

\usepackage{nameref,hyperref}

\usepackage[right]{lineno}

\usepackage{microtype}
\DisableLigatures[f]{encoding = *, family = * }

\usepackage[table]{xcolor}

\usepackage{array}
\usepackage{hhline}
\usepackage{float}

\usepackage[colorinlistoftodos]{todonotes}

\newcolumntype{+}{!{\vrule width 2pt}}

\newlength\savedwidth

\usepackage[aboveskip=1pt,labelfont=bf,labelsep=period,justification=raggedright,singlelinecheck=off]{caption}

\makeatletter
\renewcommand{\@biblabel}[1]{\quad#1.}
\makeatother

\date{}

\usepackage{comment}

\usepackage{footnote}
\makesavenoteenv{tabular}
\makesavenoteenv{table}

\usepackage{hyperref}
\usepackage{makecell}

\usepackage{subcaption}
\usepackage[labelformat=parens,labelsep=quad, skip=3pt]{caption}
\usepackage{graphicx}

\begin{document}

\vspace*{0.2in}

\begin{flushleft}
{\Large
\textbf\newline{The varying importance of extrinsic factors in the success of startup fundraising: competition at early-stage and networks at growth-stage} % Please use "sentence case" for title and headings (capitalize only the first word in a title (or heading), the first word in a subtitle (or subheading), and any proper nouns).
}
\newline

Cl\'ement Gastaud\textsuperscript{1,2},
Th\'eophile Carniel\textsuperscript{1,2},
Jean-Michel Dalle\textsuperscript{*,1,2,3}

\bigskip
\textbf{1} Agoranov, Paris, France

\textbf{2} Sorbonne Universit\'e, Paris, France

\textbf{3} i3-CNRS, \'Ecole Polytechnique, France

\bigskip

% Use the asterisk to denote corresponding authorship and provide email address in note below.
* jean-michel.dalle@sorbonne-universite.fr %(J.-M. D.)

\end{flushleft}

\section*{Abstract}
We address the issue of the factors driving startup success in raising funds. Using the popular and public startup database Crunchbase, we explicitly take into account two extrinsic characteristics of startups: the competition that the companies face, using similarity measures derived from the Word2Vec algorithm, as well as the position of investors in the investment network, pioneering the use of Graph Neural Networks (GNN), a recent deep learning technique that enables the handling of graphs as such and as a whole. We show that the different stages of fundraising, early- and growth-stage, are associated with different success factors. Our results suggest a marked relevance of startup competition for early stage while growth-stage fundraising is influenced by network features. Both of these factors tend to average out in global models, which could lead to the false impression that startup success in fundraising would mostly if not only be influenced by its intrinsic characteristics, notably those of their founders.

\section{Introduction and literature overview}
%Investors' performances are dependent on their ability to 'read' the market and its dynamics.
In the case of venture capital investments, where the risks are particularly high and startups' business models uncertain, a simple vision of the state of affairs of the startup is usually not sufficient to determine whether it will succeed or not.
Given the considerable attention that startups have drawn in the past decade from private investors and public policy makers alike, getting a better understanding of the criteria driving startup success has therefore become a particularly relevant issue.

The early literature on this topic highlighted several features of importance with respect to startup success and failure, besides earnings and assets analysis, and including human capital~\cite{bruderl1992}, founders’ psychology~\cite{rauch2007} or countries’ cultures~\cite{begley2001}, using interviews or surveys.
More recently, large-scale public databases have appeared, notably with the creation of Crunchbase~\cite{crunchbase} in 2007 that has been used in numerous recent academic studies~\cite{dalle2017}, sometimes replaced or supplemented by other similar datasets such as Dealroom~\cite{dealroom}, CBinsight~\cite{cbinsights} or Owler~\cite{owler}.

In this context, an emerging field of research has started developing predictive models based on these large-scale datasets: predicting startup success or failure~\cite{krishna2016, sharchilev2018}, predicting M\&As~\cite{xiang2012, shi2016} or predicting crowdfunding success~\cite{li2016, zhang2017}. In addition to standard economic and financial variables (funds raised, cash burn, etc.), these approaches highlight the importance of other features such as the influence of founders~\cite{krishna2016, li2016}, online and social media presence~\cite{sharchilev2018, zhang2017} or else topics associated with the company's description~\cite{yuan2016, lee2018}. According to these studies, most elements driving investors' decisions would seem to be only intrinsic to the potential investment target: its addressable market, founders, business model, etc., most notably at the expense of extrinsic, context- and environment-related features~\cite{landstrom1998}.
Some studies~\cite{sharchilev2018, dellerman2017} have started taking into consideration elements related to the environment in which the startup operates, which is particularly relevant both because of the availability of information on these environments, and because of their potential impact, notably with respect to competition or to the embeddedness of startups in given ecosystems and sectors. 
More specifically, some papers~\cite{zhang2015, liang2016} have focused on social networks, notably using distance features between investors and companies to predict investments as a link prediction task.
%While this approach can help understand the interactions between VCs and startups, and VCs between them, especially in the context of startup sourcing~\cite{fried1994}, it does not allow us to understand the business of companies. Social links is only one part of the environment surrounding a startup.
Most recently, and in the different context of predicting the long-term success of startups, Bonaventura et al.~\cite{bonaventura2019} have made use of the network of employees, in startups and other companies. Interestingly, their results show that crisis events increase unpredictability in entrepreneurial ecosystems.

However, none of these articles look at funding stages separately, although early-stage, growth and late-stage funding rounds are mostly conducted by different actors and might therefore be driven by different dynamics. Sharchilev et al.~\cite{sharchilev2018}, in order to predict future rounds, simply includes the stages of previous funding rounds as a feature but does not consider different models for each stage. To put it differently, it wouldn't be surprising at all if factors of interest for investors when choosing potential ventures would differ at the different stages of investment. By mixing all stages, some factors critical for certain rounds and not for others could be averaged out and disappear from global models.

We analyze funding stages separately in this paper. In addition, while accounting for various intrinsic characteristics of startups, we also explicitly assess the relevance of two key extrinsic features:
\begin{itemize}
    \item we explicitly account for competition by introducing dedicated metrics. Although competition is a key element suitable to affect and influence startup success, it has been mostly neglected in related works, with the notable exception of~\cite{sharchilev2018} who included two features related to competition in their model, but used metrics directly retrieved from Crunchbase though Crunchbase has notoriously been lacking in this regard, to the point that these elements have since been removed from the website. A crowdsourced database of competitors can also be found on Owler~\cite{owler}, but it is focused on large companies, which would not be appropriate to predict startup success. To the extent of our knowledge, and in the different context of predicting M\&As, only Shi et al.~\cite{shi2016} have proposed a framework to evaluate competition at a large scale by applying topic modeling techniques to startup descriptions and calculating a "business similarity" measure from this representation. Following Shi et al.~\cite{shi2016} and in accordance with the suggestions of ~\cite{dellerman2017}, we progress further in this direction by using the recent advances of natural language processing techniques to compute companies' representations.
    %\todo{tu ne m'as pas dit que dellerman aussi avait fait qqch?}
    %\todo[author=CG]{dellerman propose d'utiliser ce qu'à fait Shi et al, et il propose juste un framework sans l'implémenter. Je sais pas si ça vaut le coup d'en parler du coup.}
    \item we explicitly account for the network of startup and investor relations not only through the centrality of investors but also by pioneering on such data the use of Graph Neural Networks (GNN), a recent deep learning development~\cite{scarselli2008} that enables the handling of graphs as such, and as a whole, in machine learning prediction tasks, instead of simply using features derived from it.
\end{itemize}

We find that predicting the success of a startup when raising funds is indeed associated with different features at early and growth stages: early-stage fundraising being notably associated with competition, with a low intensity of competition markedly increasing the probability of raising funds, and with a lower relevance of network features, while growth-stage fundraising seems to be notably influenced by network features, competition playing only a limited role. Both characteristics -- competition and netwoks -- do average out in the global model, giving the (false) impression that neither competition nor networks would matter so much, compared to the characteristics of founders notably -- whereas they do matter, but at different stages of startup fundraising.

%In order to determine the importance of these features, we propose to measure their impact on future funding rounds through a prediction model: by trying to predict which startups will raise funds in the future, we learn which characteristics are most important in determining startup success.

\section{Materials and methods}
\subsection{Data used}

\begin{figure}[H]
  \begin{center}
  \includegraphics[width=14cm]{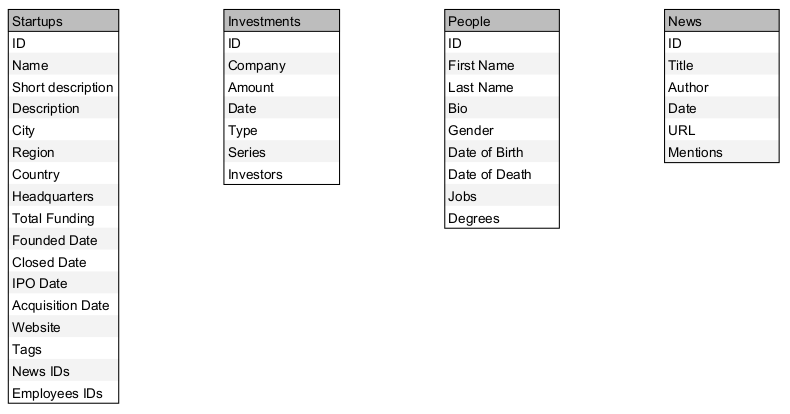}
  \caption{{\bf Data recovered from Crunchbase.}}
  \label{CrunchbaseData}
  \end{center}
\end{figure}

We exploit a dataset of startups from Crunchbase~\cite{crunchbase}, a mainstream source of data for academic research~\cite{dalle2017}. 
For each startup, we retrieved its date of creation, location, sectoral tags (describing its economic sector, technology and/or market), textual description and, most notably, all the information with respect to the funds that the startup has raised, including the date at which they were raised, the amount of funding, the nature of the funding round and the identity of the investors as well as all the articles mentioning this company available on Crunchbase (Figure~\ref{CrunchbaseData}).
In addition, we retrieved all information available about people, giving us in particular proxies with regard to the experience of startup founders.
Overall, our dataset consists in $\numprint{618366}$ companies, $\numprint{221299}$ investment rounds, $\numprint{783787}$ people and $\numprint{6363831}$ news articles.

Most of the data can be used as is in a prediction model. However, textual data that potentially contains relevant information implies preprocessing. In particular, encoding textual descriptions of startups in a vector enables similarity measurements between startups that we will use to assess competition.

\subsection{Prediction task}

It should first be noted that the definition of "success" for a startup is not straightforward. From an investor's point of view, a startup is successful if he or she is able to make a successful {\it exit}: that is, if he or she can sell his or her shares for a profit, usually through an IPO or an acquisition. Methodologically though, acquisitions are not always successful exits, as investors might not be able to recover their initial investments in this process. Furthermore, since neither entry nor exit prices are generally disclosed, it is extremely difficult to discriminate between successful and unsuccessful exits. In addition, already profitable startups might remain private for a long time: to qualify such startups as "unsuccessful" would be misleading.
Basically, apart from success in crowdfunding \cite{li2016, zhang2017} or M\&As\cite{xiang2012, shi2016}, the main option suggested in the literature to measure success is to predict future funding rounds\cite{sharchilev2018}: i.e., at a given time, determining which startups will raise funds in the next $t$ years, with $t$ a parameter.
%The reasoning is the following: if a startup raises funds, that means investors are still confident it will succeed.

In this paper, we will conduct two studies in parallel. First, for each semester between 2010 and 2015, we will study all the startups that raised funds in venture and seed rounds and try to predict if they will raise another round, get acquired or go public in the following two years. We will remove from this study startups that were already acquired or public before the end of the studied semester. Looking at the problem at different time stamps allows us to increase the number of samples, while also staying focused on \textit{active} startups.

Second, we will do the same analysis but focused on different stage fundraiser. Instead of taking all the startups that raised in a semester, we will only select those that raised in seed, series A or series B during this time.

The number of samples is reported in table \ref{samples}.

\begin{table}[ht]
\begin{center}
\begin{tabular}{|c|c|c|c|}
\hline
\textbf{Experiment} & \textbf{\# of samples} & \textbf{\# of different startups} & \textbf{\# of positive samples} \\
\hline
\textbf{All fundraisers} & 65957 & 44620 & 26697 (40\%) \\
\hline
\textbf{Seed rounds only} & 27828 & 23520 & 10338 (37\%) \\
\hline
\textbf{A rounds only} & 8564 & 7970 & 4063 (47\%) \\
\hline
\textbf{B rounds only} & 3843 & 3589 & 1892 (49\%) \\
\hline

\end{tabular}
\end{center}
\caption{Number of samples for the different experiments. Positive samples are  startups that have raised in the following 2 years. The number of startups is inferior to the number of samples because a startup can appear at different timestamps if they make several funding rounds.}
\label{samples}
\end{table}

\subsection{Investor network}
We propose that the position of a startup's investors in the investor's network is extremely important in determining its success. First, as stated in \cite{fried1994}, VCs rely heavily on their network to source startups. A startup thus has better chances of finding its next investor if its existing ones are well connected. Second, \cite{hochberg2007, stuart1999} put forward that "faced with great uncertainty about the quality of young companies, third parties rely on the prominence of the affiliates of those companies to make judgments about their quality and that young companies "endorsed" by prominent exchange partners will perform better than otherwise comparable ventures that lack prominent associates". VC reputation has also been positively linked to post-IPO performances~\cite{krishnan2011}.

In order to assess the reputation of an investor, we use its betweenness centrality in the relationship network, as it was done in \cite{hochberg2007, krishnan2011, atanasov2012}. Using Crunchbase data, we can build this graph using co-investments, investors being linked if they both invested in the same startup. Each edge of the graph is weighted, the weight being proportional to the number of portfolio companies that the investors have in common. In our prediction model, for each startup, we will use as features the max, mean and sum of the centralities of its investors.

\subsection{Competition analysis}
Assessing competition using startup descriptions requires the use of a measure of their similarity. Literature on text embedding and similarity has skyrocketed in the last few years with the rise of Natural Language Processing. Some of the most commonly used algorithms include a bag-of-words and TF-IDF model (used in Batista et al.~\cite{batista2015} to classify Crunchbase startups in categories), topic modeling approaches such as LSI or LDA~\cite{blei2003} (used in Shi et al.~\cite{shi2016}) or more recently using word embeddings such as word2vec~\cite{mikolov2013}, GloVe~\cite{pennington2014} or FastText~\cite{joulin2016}.
Among these approaches, prediction-based word-embeddings have been found to yield better results than their count-based counterparts~\cite{baroni2014}, and we will thus use a word2vec approach in this paper.

Word2vec uses a neural network to predict the most probable words around every given word. By doing so, we compute a (low-dimensional) vectorial representation of each word, encoding the "context" in which the word is used. This vectorial representation gives us the position in word-space of each word, allowing us to compute distance and similarity measures between words. The dimension of this space is a hyperparameter of the algorithm. It will be set to $h = 300$, as is commonly the case.

\paragraph{Preprocessing}
Each startup description was prepared by:
\begin{itemize}
    \item removing non-letter characters, including punctuation and numbers;
    \item lowering all characters;
    \item removing stopwords using NLTK~\cite{bird2009};
    \item removing startups which descriptions contain less than ten words;
    \item constructing bigrams (if two words are often seen together, assume they are one and only entity such as "New\_York" instead of "New" and "York". This was done using Gensim's implementation~\cite{rehurek2010}).
\end{itemize}

\begin{figure}[H]
  \begin{center}
  \includegraphics[width=14cm]{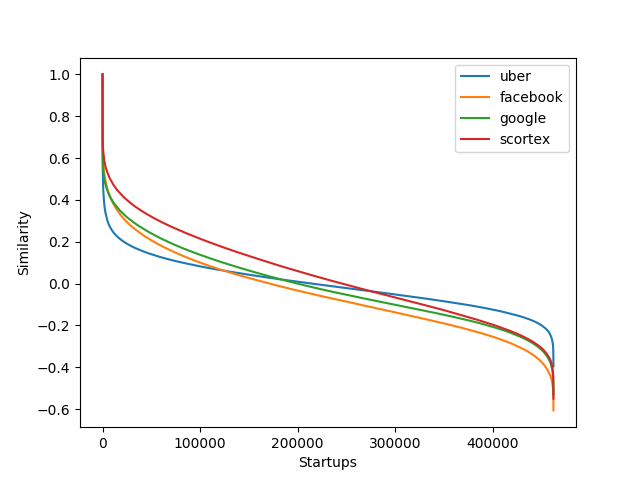}
  \caption{{\bf Similarity between some selected startups and our database (ordered by similarity)}}
  \label{Similarity}
  \end{center}
\end{figure}

\paragraph{Computing similarity}
We first learnt word embeddings using word2vec in Gensim. Then, those embeddings were used to calculate the embeddings of the descriptions following the Smooth Inverse Frequency approach proposed in \cite{arora2016}.
The similarity between two embeddings was measured using standard cosine similarity: given two vectors $A$ and $B$, the similarity between them is:
\begin{equation}
    similarity = \frac{A \cdot B}{||A|| \cdot ||B||}
\end{equation}
Two organizations were deemed "in competition" if their similarity was superior to a given threshold parameter $min\_sim$. Figure \ref{Similarity} shows the similarity measure of some selected startups against the whole database.
We can see that the similarity decreases rapidly, as expected.
$min\_sim$ was set at $0.5$: this value was chosen arbitrarily so that it was high enough to ensure similarity of competing companies, but low enough to ensure the identification of competitors most of the time.

For each startup, we then obtain a list of similar companies that we approximate as its competitors, and from which we derive three features: the number of competitors, funds raised by competitors, and funds raised in the last year by these competitors.

\subsection{Prediction Model}
\label{predictionModel}
All features used in our models are presented table~\ref{features}.
Data was scaled using scikit-learn's~\cite{pedregosa2011} implementation of the Yeo-Johnson's transformation~\cite{yeo2000}. This transformation improves the normality of our features, which is especially useful in order to make sense of highly skewed distributions, as it is the case with funding amounts for instance in our case.

Out of the most standard classification algorithms, random forest was the most proficient after fine-tuning. we will use this algorithm as the baseline for the prediction task.

Metrics used to evaluate each model are: macro-average's precision, recall and f1-score, ROC-AUC and the precision of the $n$ most probable positive samples (noted $P@n$). Results presented here are mean values of five runs of each algorithm.

\begin{table}[ht]
\begin{center}
\begin{tabular}{|c|c|}
\hline
\textbf{Feature} & \textbf{Description} \\
\hline
\# startups & Number of startups in the region (as defined by Crunchbase) \\
\hline
Founded date & Time since the creation of the startups in days \\
\hline
Total Funding & Total funding raised up to this date in dollars\footnote{Funding rounds' amounts were often undisclosed. Those amounts were estimated as the median amount of all the funding rounds of the same type and same round.} \\
\hline
Number of Rounds & Number of funding rounds up to this date \\
\hline
Last Round Date & Time since last funding round in days \\
\hline
Last Round Funding & Funding raised in last round in dollars \\
\hline
First round Date & Time since first funding round in days \\
\hline
Has Seed/A/B/C/D & True if the startup has raised in Seed/A/B/C/D (5 different features) \\
\hline
\# of investors & Number of investors in previous rounds \\
\hline
Max investor centrality & maximal betweenness centrality of the startup's investors \\
\hline
Sum investor centrality & Sum of the betweenness centrality of the startup's investors \\
\hline
Max Portfolio & Maximal size of the startup's investors' portfolios \\
\hline
Nb articles & Number of articles on Crunchbase mentioning the startup \\
\hline
News Increase & \makecell{Year on year augmentation of the number of articles \\ on Crunchbase mentioning the startup} \\
\hline
\# founders & Number of founders reported on Crunchbase \\
\hline
\# previous startups & Number of previous startups created by the founders before this one \\
\hline
Competition startups & Number of startups identified as competitors \\
\hline
Competition funding & Money raised by competitors in dollars \\
\hline
Competition funding 1y & Money raised by competitors in the last year in dollars \\
\hline

\end{tabular}
\end{center}
\caption{Features used in the prediction model}
\label{features}
\end{table}

\subsection{Introducing Graph Neural Networks}
Recent advancements in machine learning and neural networks have proposed new ways of analyzing data organized in graph forms. From vanilla graph neural networks~\cite{scarselli2008} to convolutional~\cite{kipf2016} and attention~\cite{velickovic2017} variants, this approach has attracted a lot of attention especially in the past couple years.
Given the startup-investor structure of our data, we propose to apply such models so as to capture information about the organization and position of startups in the entrepreneurial ecosystem. Predicting successful startups can be reframed as a binary node classification problem, where we want to predict the label of a subset of the nodes. 
More specifically, we will consider here a bipartite graph between startups and investors.
One advantage of this approach is that we can take into consideration all the startups we have information about, and not only the ones we want to predict the label of: in the case of exited companies for instance, trying to predict its next funding round does not make much sense. However, taking it in consideration in the context of a graph can help us predict future funding rounds of neighboring startups.

We will focus this study on Graph Convolution Networks~\cite{kipf2016} (GCN). Simply put, given a list of node attributes and a list of edges, we teach a neural network to propagate information along edges, obtaining for each node a probability of belonging to a certain class (in our case, raising funds or not in the next $t$ years). Just like convolutional layers learn how to filter information of neighboring pixels in an image, we learn here how to filter information from neighboring nodes in the graph.

We will construct our graph as follows: for each round of each startup, a link will be drawn between the startup and the investors of this round. If a company raise funds with the same investor at different stages, several edges will connect the two nodes in a multigraph fashion. For ease of use, companies that also invest in other startups will be split into entrepreneurial and investing entities. Self-loops were also added for each node following the renormalization trick introduced in \cite{kipf2016}, in order to update each node's information with its own features instead of simply the features of its neighbors.
For each node in this graph, we indicate if its an investor or a startup and all the features mentioned in the last section. In the case of investors, these features are all set to zero.

We used a two-layer GCN in this study, with a hidden dimension of 64.
The model is given:
\begin{itemize}
    \item the adjacency matrix $A$ of the graph,
    \item A list of attributes for each nodes $X$ of size $N \times D$, with $N$ the number of nodes and $D$ the number of features
\end{itemize}
The global formulation of this neural network is then:
\begin{equation}
    Z = softmax(\hat{A} ReLU(\hat{A} X W^{(0)})W^{(1)})
\end{equation}
with $Z$ the output for each node, $W^{(n)}$ the weight matrix of the $n$-th layer and $\hat{A} = D^{-\frac{1}{2}} A D^{-\frac{1}{2}}$ with $D$ the diagonal node degree matrix of $A$.

We can note that this architecture produce an output for each node, even startups for which we are not trying to predict a funding round as well as investors. These outputs will simply not be taken into consideration in the training, and is just an artefact of our model.

\section{Results}

\subsection{Random Forest}
When compiling all data from 2010 to 2015, we get a dataset of $\numprint{65957}$ samples with $\numprint{26697}$ (40\%) positive examples (see Table \ref{samples}). For the entire dataset, results are reasonably higher than random (cf figure \ref{results}), with a ROC-AUC of 63.3\% and a precision of 80.4\% for the top 100 predictions, but removing competition features does not decrease drastically the quality of the prediction except for top 50 or 100 predictions. When we focus on post-seed prediction, which corresponds to a dataset of $\numprint{27828}$ samples, $\numprint{10338}$ of which were positive (37\%), results are markedly higher than random and competition appears as a critical feature, since removing it reduces the f1-score by 6 percentage points, and even more when looking at the precision of the 100 most probable successful startups since this precision goes from 66.8 to 83.2\% of success without and with competition features, respectively.
For later funding rounds (Series A and B), we find that the influence of competition decreases if not disappears, in a context however where overall performance also drops notably due to a reduced dataset. Altogether, these results suggest a marked effect of competitive features mostly with respect to post-seed fundraising.

\begin{table}[ht]
\begin{center}
\begin{tabular}{|c|c|c|c|c|c|c|}
\hline
\textbf{Model} & \textbf{ROC-AUC} & \textbf{Precision} & \textbf{Recall} & \textbf{F1-score} & \textbf{P@50} & \textbf{P@100} \\
\hline
\textbf{All} & 63.3 & 63 & 63 & 63 & 82.4 & 80.4 \\
\hline
\textbf{All w/o comp} & 62.3 & 62 & 62 & 62 & 70.0 & 72.6 \\
\hline
\textbf{Seed} & 67.8 & 69 & 68 & 68 & 87.6 & 83.2 \\
\hline
\textbf{Seed w/o comp} & 62.3 & 62 & 62 & 62 & 69.2 & 66.8 \\
\hline
\textbf{Series A} & 63.2 & 63 & 63 & 63 & 76.8 & 73.0 \\
\hline
\textbf{Series A w/o comp} & 61.9 & 62 & 62 & 62 & 68.4 & 70.0 \\
\hline
\textbf{Series B} & 60.6 & 61 & 61 & 61 & 67.2 & 64.0 \\
\hline
\textbf{Series B w/o comp} & 60.3 & 60 & 60 & 60 & 65.2 & 61.8 \\
\hline

\end{tabular}
\end{center}
\caption{Results of the random forest prediction models in percentage for different rounds, with or without the competition features. Values are the means of five runs.}
\label{results}
\end{table}

\begin{figure}
  \centering
  \begin{subfigure}{8.5cm}
    \centering\includegraphics[width=8cm]{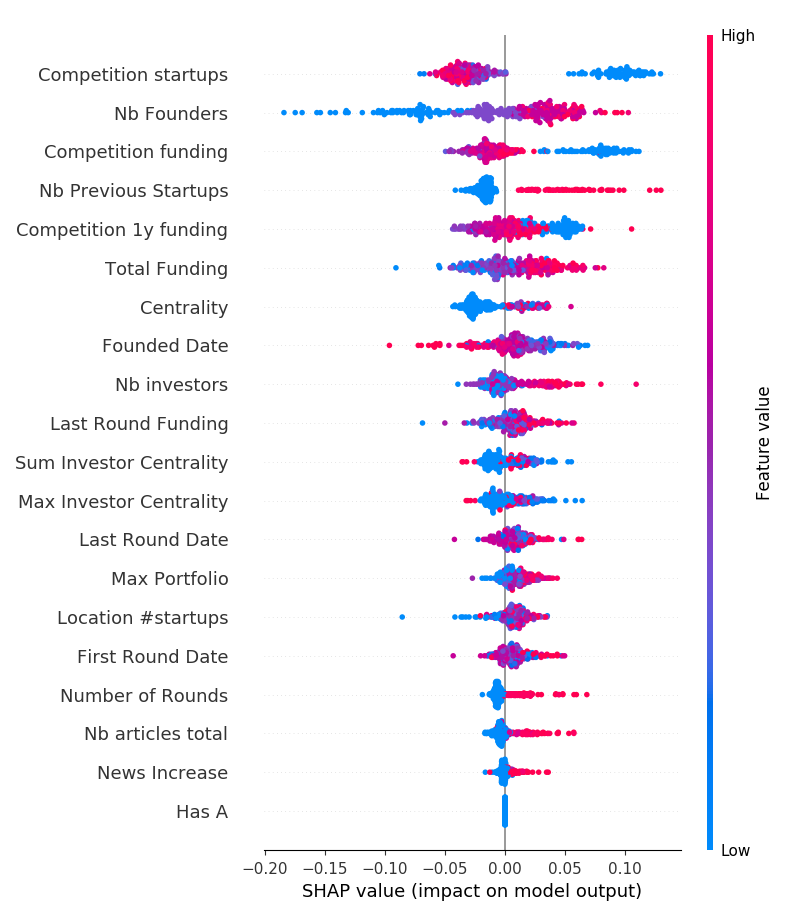}
    \caption{Post-seed rounds}
  \end{subfigure}%
  \begin{subfigure}{8.5cm}
    \centering\includegraphics[width=8cm]{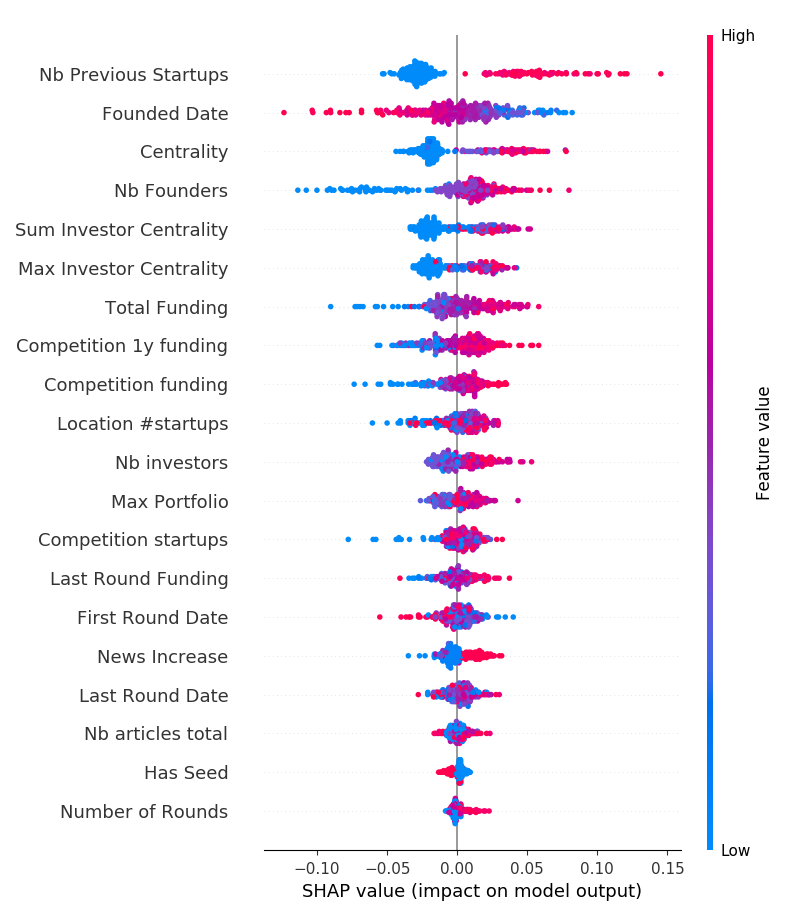}
    \caption{Post-A rounds}
  \end{subfigure}\vspace{10pt}

  \begin{subfigure}{8.5cm}
    \centering\includegraphics[width=8cm]{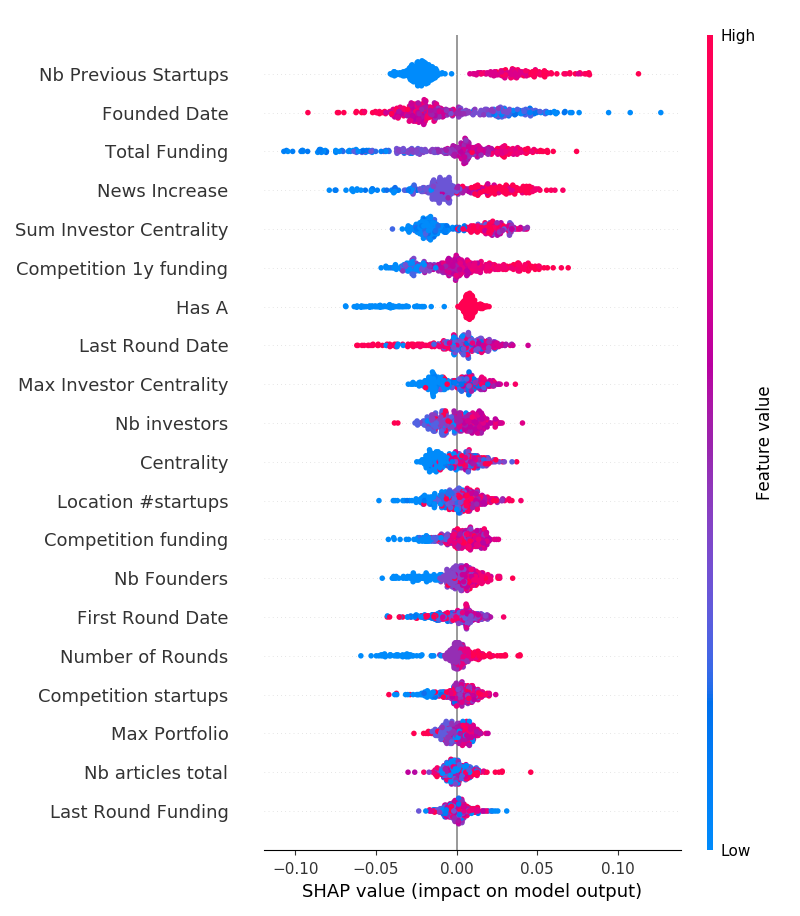}
    \caption{Post-B rounds}
  \end{subfigure}%
  \begin{subfigure}{8.5cm}
    \centering\includegraphics[width=8cm]{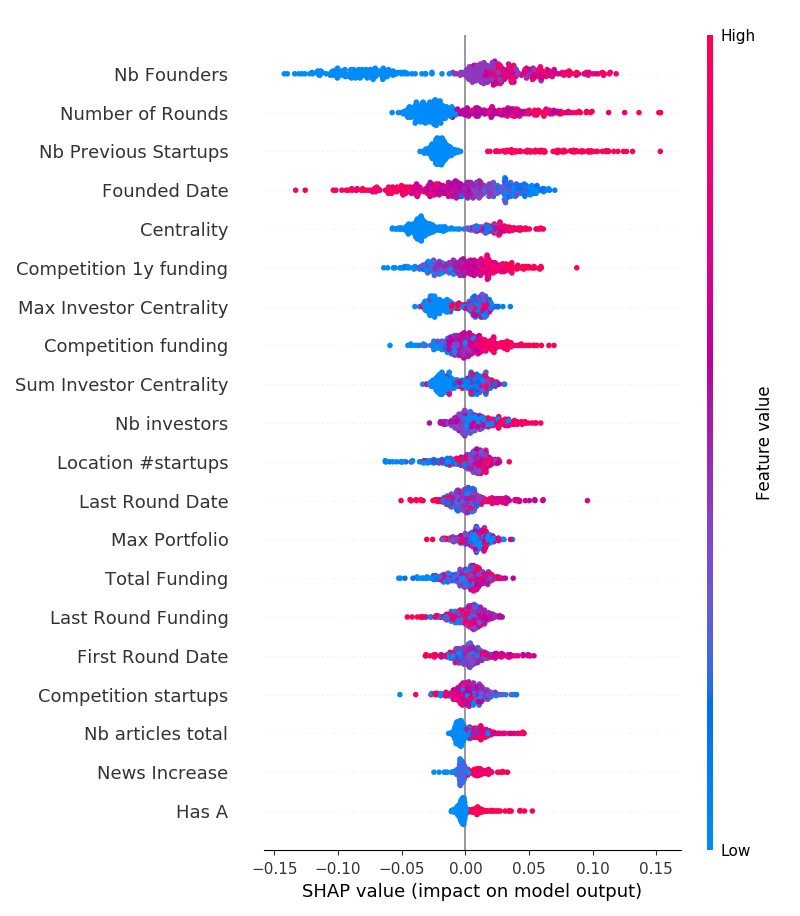}
    \caption{All rounds}
  \end{subfigure}
  \caption{{\bf Importance of all features in our classification model.} These plot shows the impact of the values of the different features on the output of the prediction model.}
  \label{featureImportance}
\end{figure}

%\subsection{Feature importance}
For all experiments, we computed the SHAP values~\cite{lundberg2018} of the features in order to compare their impact (cf figure~\ref{featureImportance}). This analysis comforts our previous observation that competition is a major feature in post-seed prediction, but this effect decreases with stages.
In addition this analysis shows that competition in general is negatively correlated to the probability of a post-seed funding round, while the funding of competitors in the last year is \textit{positively} correlated. Inversely, competition is a positive feature in post-A, post-B and when considering all funding rounds, although less intensely.

Altogether, these results suggest that post-seed fund raising is strongly affected by the competition a startup faces. Too intense competition seems to impact negatively the prospects of a startup raising post-seed funds, which is understandable if it signals the fact that a startup is only the $n$-th attempts to address an economic issue for which many others have already raised funds. Furthermore, recent competitors' funding is a good indicator that the startup is tackling a problem that is still ongoing.

Centrality, on the other hand, exhibit a more timid impact on model output especially for early stage funding. For later stages, the effect is clearer, centrality of the investors being a positive influence on our prediction model.
We thus postulate that in early stage, companies are still looking for their market which makes competition an excellent proxy of their ability to do so. In later stages, startups should have found their market if there is one, and this criteria diminishes in importance. At the same time, startups have by then raised sufficiently that the identity of their investors sends a strong signal about the quality of the project, and investor network-based metrics become important in the decision-making process of future investors.

This plot also confirms previous intuitions: for example, founder experience, measured as the number of previous companies, is indeed a factor positively correlated to funding rounds for all investment stages.
The number of founders also appears to be a crucial factor in assessing a startup's potential, a fact observed in previous studies such as \cite{eisenhardt1990}. Indeed, smaller teams, especially solo founders, are less likely to master all the skills necessary to successfully make their company grow and are thus shunned by investors.
It should be noted nonetheless that Crunchbase is not exhaustive in listing the founders, and a bias might exist where successful startups are more likely to have their founders accounted for in the database.

\subsection{Graph neural networks}
We further study the prediction of startup success in raising funds using GNNs, in order to better analyze the influence of networks on the predictions.
Globally, the results with this framework (including competition features) are mostly similar to the previous ones, with the notable disappearance of the increase in accuracy for post-seed rounds~\ref{resultsGNN}, which is fully in line with the previous finding according to which networks would be of a lesser relevance for post-seed rounds, compared to later ones. The precision of the prediction of the most probable successful startups in terms of raising funds also increases considerably, which could be of peculiar relevance for investors. Again, this is consistent with the fact that, since reputation becomes more important than intrinsic features for highly probable events~\cite{stuart1999}, a network-first approach would be associated with better results for such metrics. Finally, with respect to the performance of the GNN framework more generally, and in relation to the fact that results for B rounds were not consistent and are not reported here, it should be noted that a propagation model such as GNNs is disadvantaged by the fact that only a small fraction of the nodes in the graph is actually "helpful" in the prediction task, most specially at later stages: indeed, very few startups raise funds in B compared to the total number of startups.
%We impute the decrease in performance for stage-by-stage experiments to the relatively low number of sample, which is crucial in such a framework, and in deep learning frameworks in general. In particular since , 

\begin{table}[ht]
\begin{center}
\begin{tabular}{|c|c|c|c|c|c|c|}
\hline
\textbf{Model} & \textbf{ROC-AUC} & \textbf{Precision} & \textbf{Recall} & \textbf{F1-score} & \textbf{P@50} & \textbf{P@100} \\
\hline
\textbf{All} & 63.2 & 65 & 63 & 63 & 90.4 & 85.6 \\
\hline
\textbf{Seed} & 64.3 & 64 & 64 & 64 & 80.4 & 74.8 \\
\hline
\textbf{Series A} & 61.5 & 63 & 62 & 59.6 & 86.0 & 78.0 \\
\hline

\end{tabular}
\end{center}
\caption{Results of the GNN model in percentage for different rounds. Values are the means of five runs.}
\label{resultsGNN}
\end{table}

\section{Conclusion}
In this paper, we have presented a framework that makes use of a prediction model to analyze the factors influencing startup funding rounds. We have assessed in this context the relevance of two important extrinsic characteristics of startups, namely competition and investment networks. To do so, we have implemented algorithms so as to estimate the competition faced by a startup based on the similarity of startup descriptions and we have used it in machine learning frameworks, including the recent Graph Convolutional Networks. We found that the effect of competition on fundraising seems particularly strong for post-seed, so-called early stage, fundraising, while networks are more meaningful for later, growth stages. On the contrary, the relevance of these extrinsic factors appears confounded in more general models that span several funding rounds. Altogether, these results, that definitely deserve further investigations, suggest dynamics where being late markedly decreases the odds of being funded at early stage, and where networks play a stronger role later on.

\newpage
\bibliographystyle{unsrtnat}

\end{document}